\documentclass{article}
\usepackage{spconf,amsmath,amssymb,graphicx,booktabs,multirow,hyperref}
\usepackage{tikz}
\usetikzlibrary{positioning,arrows.meta}
\hypersetup{hidelinks}
\newcommand{\method}{Speech-Rewarded Style Planning}
\newcommand{\model}{SRSP}

\title{BEYOND SPEECH CAPTIONS: SPEECH-REWARDED STYLE PLANNING FOR CONVERSATIONAL TEXT-TO-SPEECH}

\name{
Shiao Zhu$^{1}$,
Lianbo Liu$^{2}$,
Sizhen Lyu$^{1}$,
Yuzhe Wang$^{1}$,
Sheng Li$^{1}$,
Takahiro Shinozaki$^{1}$
}
\address{
$^{1}$Institute of Science Tokyo, Tokyo, Japan\\
$^{2}$Independent Researcher, Tokyo, Japan  
}
\ninept
\begin{document}
\maketitle

\begin{abstract}
Natural-language style descriptions provide an interpretable interface
between large language models (LLMs) and controllable text-to-speech
(TTS). However, using descriptions as pseudo-labels compresses target
acoustics into text, and descriptive fidelity need not imply effective
control of a particular synthesizer. We empirically show that speech--text
alignment only weakly predicts downstream acoustic similarity among
candidate instructions for the same utterance. We therefore propose
\method{} (\model{}), which trains a text-based style planner through a
frozen downstream TTS model. Given dialogue history and response text, the planner generates candidate instructions and is optimized with group-relative policy
optimization (GRPO), using the teacher-forced likelihood of target
speech tokens as the reward. On an English subset of the ISCSLP 2026
CoT-TTS corpus, \model{} achieves higher speech-style and emotion
similarity to target speech and lower mel-cepstral distortion than the
Base LLM and target-audio-informed captioning baselines. LLM-based
expressive speech evaluation further shows gains over all baselines in
contextual appropriateness and reference consistency.
\end{abstract}

\begin{keywords}
conversational speech synthesis, speaking style control, large language models, reinforcement learning, spoken dialogue systems
\end{keywords}

\section{Introduction}
\label{sec:intro}

Spoken dialogue conveys both linguistic content and manner of delivery.
The same response can sound reassuring, hesitant, or excited, with
different conversational effects. Response text alone does not fully
determine how an utterance should be delivered. Inferring an appropriate
speaking style from conversational context is therefore an important task
in conversational TTS \cite{lin2024spokenllm,yan2024perceptiveagent}.

Natural-language style descriptions provide an interpretable and editable
interface between conversational reasoning and prompt-based TTS,
allowing textual instructions to control how an utterance is spoken
\cite{lyth2024natural,yu2025implicit,lou2025parastyletts}. Speech captioners such as StyleCap, SECap,
and AlignCap describe paralinguistic and emotional properties of speech
in natural language
\cite{yamauchi2024stylecap,xu2024secap,liang2024aligncap}.
Such descriptions have also been used as explicit intermediate
representations in conversational speech generation; for example,
Chain-Talker predicts a context-aware empathetic caption before
generating semantic speech codes and expressive speech
\cite{hu2025chaintalker}. A natural way to train a standalone
text-based style planner is therefore to use speech-derived captions
as supervision for predicting style instructions from dialogue context
and response text.

However, when target speech is summarized as a caption, acoustic details
omitted from that summary are not explicitly represented in the
caption-based supervision. More importantly, predicting a target
description and producing an effective instruction for a particular
synthesizer are distinct objectives: even an accurate description
need not induce the closest acoustic realization of the target speech.
We address this mismatch with \method{} (\model{}), which uses a frozen
downstream TTS model to provide training feedback for a text-based style
planner. GRPO updates the planner using rewards based on the target-speech
likelihood induced by candidate instructions, optimizing it for downstream
synthesis behavior rather than agreement with speech-derived descriptions.
Related work has also explored target-speech likelihood rewards for
expressive TTS \cite{ren2026mclp} and GRPO for instruction-controlled
TTS \cite{chen2026flexivoice}.

Our contributions are as follows:
\begin{itemize}
    \item We empirically show that, among candidate style instructions for the same utterance, speech--text alignment only weakly predicts downstream acoustic similarity, making it a weak proxy for selecting effective instructions.

    \item We propose \method{}, a text-based style planner trained with
    direct feedback from a frozen downstream TTS model. Candidate
    instructions are rewarded using teacher-forced target-speech
    likelihood, and GRPO updates the planner without requiring target
    style captions.

    \item Experiments on an English subset of the ISCSLP 2026 CoT-TTS corpus
show that \model{} outperforms all baselines on five downstream
acoustic metrics across both evaluation partitions. LLM-based expressive
speech evaluation further shows gains in contextual appropriateness
and reference consistency.
\end{itemize}

\section{Descriptive Alignment as a Proxy for Downstream Control}
\label{sec:description-analysis}

\begin{figure*}[t]
  \centering
  \includegraphics[width=.92\textwidth]{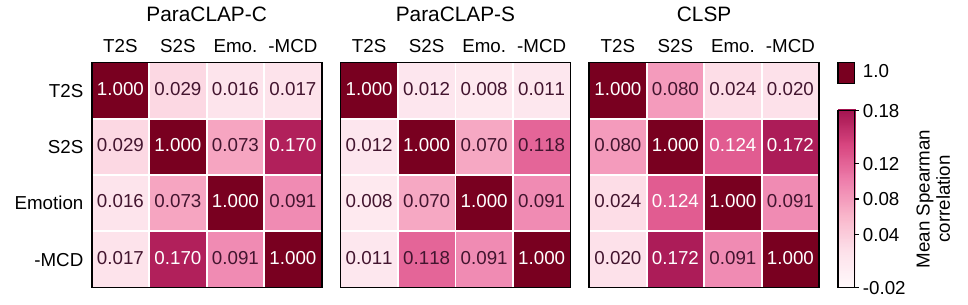}
  \caption{Mean within-utterance Spearman correlations across both development partitions.}
  \label{fig:candidate-correlation}
\end{figure*}

We examine whether stronger speech--text alignment of a style instruction
implies better downstream control of the synthesizer. For each utterance
in the 5,500-utterance development set described in
Sec.~\ref{sec:setup}, we sample eight candidate style instructions from
the Base LLM given the dialogue history and response text, using the
training decoding parameters. We then synthesize the response under
each instruction.

We use CLAPScore \cite{elizalde2023clap} to measure style alignment between the generated style
instruction and target speech, denoted as T2S, and speech--speech cosine
similarity to measure style similarity between synthesized and target
speech, denoted as S2S. Both are computed as cosine similarities between
embeddings from pretrained speech--text dual-encoder models. Specifically, we
use ParaSpeechCLAP-Combined (ParaCLAP-C), ParaSpeechCLAP-Situational
(ParaCLAP-S) \cite{diwan2026paraspeechclap}, and CLSP
\cite{yang2026clsp}. Together with the three S2S scores, we use emotion
similarity, computed as the cosine similarity between emotion2vec
probability vectors \cite{ma2024emotion2vec}, and MCD-DTW, the
mel-cepstral distortion after dynamic time warping, as downstream
acoustic metrics.

To assess how T2S relates to downstream acoustic similarity, we compute
within-utterance Spearman correlations across candidate instructions
between each T2S score and each downstream acoustic metric, as well as
pairwise correlations among the acoustic metrics. MCD is negated so that
higher values consistently indicate better target matching, and the
correlations are averaged across utterances.

Figure~\ref{fig:candidate-correlation} shows that T2S scores from both
ParaSpeechCLAP variants are nearly uncorrelated with S2S, emotion
similarity, and $-\mathrm{MCD}$. CLSP, which targets fine-grained
speech-style alignment \cite{yang2026clsp}, shows somewhat stronger
but still weak T2S correlations. For each downstream acoustic metric,
its correlation with T2S remains lower than its correlations with
the other acoustic metrics.

To examine the downstream consequences of this ranking mismatch,
we compare instruction selection based on T2S with selection based on
downstream acoustic metrics. For each utterance, we select the best-scoring
candidate under each criterion and evaluate the corresponding synthesis
on all five downstream acoustic metrics, using random selection as
a reference.

\begin{table}[t]
\centering
\caption{Synthesis performance by instruction-selection criterion on
the combined development partitions. Bold and underline indicate the
best and second-best results, respectively.}
\label{tab:description-selection}
\small
\setlength{\tabcolsep}{2.2pt}
\newcommand{\selectionresult}[3]{#1/#2/#3}
\begin{tabular}{lccc}
\toprule
Selection & S2S C/S/CLSP $\uparrow$ & Emo. $\uparrow$ & MCD $\downarrow$ \\
\midrule
Random
& \selectionresult{.3105}{.3049}{.8135} & .5964 & 6.3266 \\
ParaCLAP-C T2S
& \selectionresult{.3181}{.3127}{.8180} & .6044 & 6.3052 \\
ParaCLAP-S T2S
& \selectionresult{.3125}{.3104}{.8163} & .5986 & 6.3142 \\
CLSP T2S
& \selectionresult{.3173}{.3101}{.8198} & .6032 & 6.2711 \\
\midrule
ParaCLAP-C S2S
& \selectionresult{\textbf{.5133}}{\underline{.4595}}{\underline{.8465}}
& .6210 & 6.0731 \\
ParaCLAP-S S2S
& \selectionresult{\underline{.4529}}{\textbf{.5262}}{.8406}
& .6176 & 6.1581 \\
CLSP S2S
& \selectionresult{.4200}{.4075}{\textbf{.8762}}
& \underline{.6354} & \underline{6.0677} \\
Emotion
& \selectionresult{.3247}{.3191}{.8206} & \textbf{.8601} & 6.1932 \\
MCD-DTW
& \selectionresult{.3456}{.3288}{.8235} & .6160 & \textbf{5.2331} \\
\bottomrule
\end{tabular}
\end{table}

Table~\ref{tab:description-selection} shows that T2S-based selection
provides only small improvements over random selection. In contrast, selecting by any downstream acoustic metric outperforms
all three T2S-based criteria across all five downstream acoustic metrics,
including those not used for selection. Thus, the candidate pool contains instructions
that produce speech more closely matching the target, but descriptive
alignment does not reliably identify them. Together with the weak
correlations in Fig.~\ref{fig:candidate-correlation}, these results
indicate that descriptive alignment is a weak proxy for downstream
synthesis behavior, motivating synthesis-aware feedback.
\section{Speech-Rewarded Style Planning}
\label{sec:method}

\subsection{Problem formulation}

Let $H$ be text-only dialogue history, $y$ the response text, $p$ a
same-speaker prompt utterance, and $s=(s_1,\ldots,s_T)$ the target speech
tokens. The style policy generates a natural-language instruction
$z\sim\pi_\theta(z\mid H,y)$. During training, $s$ is used only to construct
the reward. At inference, the planner receives only $(H,y)$, and its
instruction conditions the TTS model together with response text $y$ and
prompt audio $p$.

\subsection{Speech-token likelihood as reward}

For candidate instruction $z_i$, the autoregressive LLM of the frozen TTS
model computes the mean teacher-forced cross-entropy (CE) loss over the
target speech-token sequence:
\begin{equation}
\mathcal{L}_{\mathrm{TTS}}(z_i)=
-\frac{1}{T}\sum_{t=1}^{T}
\log p_\phi(s_t\mid s_{<t},y,z_i).
\label{eq:tts-loss}
\end{equation}
All candidates in a reward group share $(y,s)$, so their relative
losses measure how each instruction changes the probability assigned to
the same target speech by the fixed synthesizer. 
\begin{figure*}[t]
    \centering
    \includegraphics[width=\textwidth]{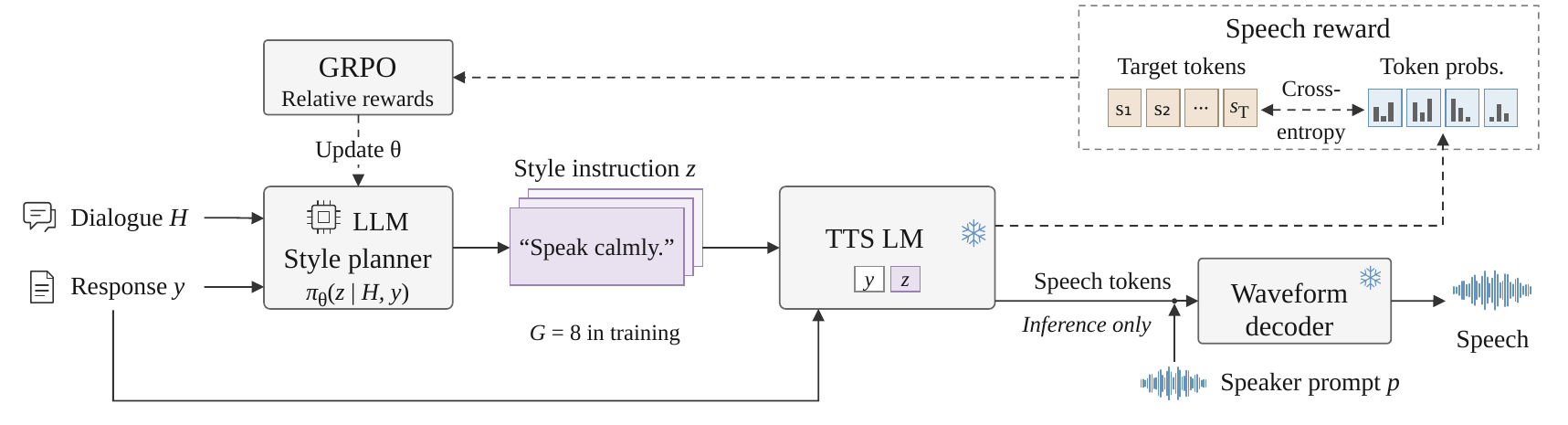}
    \caption{Overview of \model{}. Frozen-TTS target-speech likelihood rewards candidate instructions during training, while GRPO updates the planner. Dashed arrows denote training-only paths; snowflakes denote frozen components.}

    \label{fig:method}
\end{figure*}
\begin{table*}[t]
\centering
\caption{Results on \texttt{test\_in}/\texttt{test\_out}. Bold and underline indicate the best and second-best results, respectively. $^\dagger$: descriptions conditioned on the ground-truth target waveform. Paired Wilcoxon tests with Holm correction across 40 downstream acoustic
comparisons show significant gains ($p_{\mathrm{Holm}}<.05$) in 38 cases;
the exceptions are \texttt{test\_out} emotion similarity vs. Raw TTS and
Qwen3-Omni.}

\label{tab:main}
\small
\setlength{\tabcolsep}{2.5pt}
\newcommand{\mainresult}[2]{#1/#2}
\begin{tabular}{lcccccccc}
\toprule
\multirow{2}{*}{System}
& \multicolumn{3}{c}{T2S Score $\uparrow$}
& \multicolumn{3}{c}{S2S Score $\uparrow$}
& \multirow{2}{*}{Emo. Cos. $\uparrow$}
& \multirow{2}{*}{MCD-DTW $\downarrow$} \\
\cmidrule(lr){2-4}
\cmidrule(lr){5-7}
& ParaCLAP-C & ParaCLAP-S & CLSP
& ParaCLAP-C & ParaCLAP-S & CLSP
& & \\
\midrule
Raw TTS
& -- & -- & --
& \mainresult{\underline{.3370}}{\underline{.3494}}
& \mainresult{\underline{.3246}}{\underline{.3303}}
& \mainresult{.8160}{.8148}
& \mainresult{.6098}{.6119}
& \mainresult{6.2426}{6.3932} \\
Base LLM
& \mainresult{-.0089}{-.0141}
& \mainresult{.0161}{.0160}
& \mainresult{.4718}{.4686}
& \mainresult{.3166}{.3220}
& \mainresult{.2979}{.3078}
& \mainresult{.8150}{.8146}
& \mainresult{.6124}{.5927}
& \mainresult{6.3395}{6.5072} \\
AF-Next$^\dagger$
& \mainresult{\textbf{.0788}}{\textbf{.0841}}
& \mainresult{\textbf{.0996}}{\textbf{.0973}}
& \mainresult{\textbf{.5404}}{\textbf{.5423}}
& \mainresult{.2874}{.2982}
& \mainresult{.2596}{.2637}
& \mainresult{.7842}{.7863}
& \mainresult{.5770}{.5677}
& \mainresult{6.8890}{7.0128} \\
Qwen3-Omni$^\dagger$
& \mainresult{-.0087}{-.0119}
& \mainresult{.0225}{.0193}
& \mainresult{\underline{.5101}}{\underline{.5053}}
& \mainresult{.3176}{.3312}
& \mainresult{.3083}{.3130}
& \mainresult{\underline{.8194}}{\underline{.8182}}
& \mainresult{\underline{.6228}}{\underline{.6226}}
& \mainresult{\underline{6.1593}}{\underline{6.3189}} \\
\midrule
\model{} (ours)
& \mainresult{\underline{.0073}}{\underline{-.0044}}
& \mainresult{\underline{.0324}}{\underline{.0232}}
& \mainresult{.4591}{.4518}
& \mainresult{\textbf{.3632}}{\textbf{.3627}}
& \mainresult{\textbf{.3496}}{\textbf{.3502}}
& \mainresult{\textbf{.8329}}{\textbf{.8296}}
& \mainresult{\textbf{.6417}}{\textbf{.6269}}
& \mainresult{\textbf{5.9569}}{\textbf{6.1637}} \\
\bottomrule
\end{tabular}
\end{table*}
The reward is
$R_i=-\mathcal{L}_{\mathrm{TTS}}(z_i)-\mathcal{P}_i$,
where $\mathcal{P}_i$ contains auxiliary penalties that discourage
degenerate style instructions, including overly short, generic, and
repetitive outputs.

\subsection{Group-relative policy optimization}

We optimize the planner using GRPO \cite{shao2024deepseekmath}.
For each training instance, we sample $G$ instructions and standardize
their rewards within the group:
\begin{equation}
A_i=\frac{R_i-\bar{R}}{\sigma_R+\epsilon},
\end{equation}
where $\bar{R}$ and $\sigma_R$ are the group mean and standard deviation,
and $\epsilon$ is a small constant for numerical stability.

The clipped  objective includes a KL
penalty to the reference policy:
\begin{equation}
\begin{aligned}
\mathcal{J}(\theta)
=\mathbb{E}_{i,t}\Big[&
\min\{r_{i,t}A_i,
\operatorname{clip}(r_{i,t},1-\eta,1+\eta)A_i\}\\
&-\beta D_{\mathrm{KL}}(\pi_\theta\Vert\pi_{\mathrm{ref}})
\Big],
\end{aligned}
\label{eq:grpo}
\end{equation}
where $r_{i,t}$ is the token-level likelihood ratio to the rollout
policy, the expectation averages over valid generated tokens in
each rollout batch, and $\pi_{\mathrm{ref}}$ is the frozen initial
policy. The clipping parameter is $\eta$, and $\beta$ is the KL weight.

The planner is prompted to generate one natural-language English sentence
describing the speaking style, as summarized in
Fig.~\ref{fig:prompts}(a). During training, we randomly sample zero to three CosyVoice-style examples as in-context demonstrations for each instance; the same examples are shared by all candidates in a reward group. At inference, two examples are sampled per utterance.

\section{Experiments}
\label{sec:experiments}

\subsection{Experimental setup}
\label{sec:setup}
\begin{figure*}[t]
    \centering
    \includegraphics[width=\textwidth]{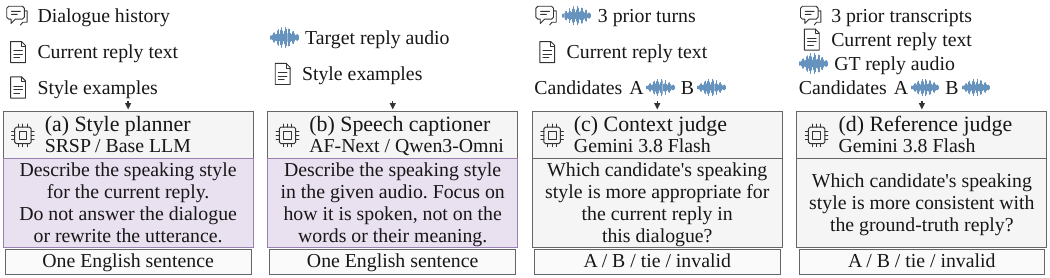}
    \caption{Abbreviated prompting configurations for style generation
and LLM-based expressive speech evaluation.}
    \label{fig:prompts}
\end{figure*}
\textbf{Data.}
We construct an English subset of the ISCSLP 2026 CoT-TTS
corpus \cite{xue2026cottts}, retaining utterances with duration
3--15\,s, active-speech ratio $\geq 0.70$, loudness
$\geq -40$\,dBFS, transcript length $\leq 300$ characters,
reference similarity $\geq 0.65$, naturalness score $\geq 3.0$,
and noise score $\geq 3.5$.
Excluding samples with temporally overlapping context and target
speech yields a 299.75-hour pool.
From this pool, we create four development and test partitions,
each containing 2,750 utterances (approximately 5 hours).
The \texttt{dev\_in}/\texttt{test\_in} partitions hold out complete
scenes from movies represented in training, whereas
\texttt{dev\_out}/\texttt{test\_out} hold out entire movies.
After excluding training samples that overlap with held-out development or test utterances in either the target or dialogue context, 277.60 hours remain for training.

\textbf{Models and optimization.}
The Qwen3.5-9B planner \cite{qwen2026qwen35} is adapted with rank-16 LoRA
\cite{hu2022lora} on all linear layers ($\alpha=32$, dropout $0.02$).
Fun-CosyVoice3-0.5B \cite{du2025cosyvoice3} is the frozen reward and
synthesis model. We train for one epoch with $G=8$, learning rate
$2\times10^{-6}$, clipping parameter $\eta=0.2$, and KL weight
$\beta=0.075$.
We use TRL 1.3.0 \cite{vonwerra2020trl} with its default token-level importance
sampling and DAPO-style loss normalization \cite{yu2025dapo}.
Style generation uses temperature $1.3$ during training and $0.7$
at test time.
For the auxiliary reward penalties, instructions shorter than 6 words receive a penalty of 5, while those containing 6--11 words receive a penalty of 0.5. Repeated bigrams and trigrams are penalized with weight $0.01$ using a per-worker rolling history of 2,048 generated instructions.

\textbf{Baselines.}
\emph{Raw TTS} synthesizes responses without inferred style instructions.
\emph{Base LLM} uses unadapted Qwen3.5-9B with the same $(H,y)$ input,
planner prompt, demonstrations, and inference configuration as \model{},
providing the direct pre-RL comparison.
\emph{AF-Next} uses NVIDIA's
\href{https://huggingface.co/nvidia/audio-flamingo-next-captioner-hf}
{Audio Flamingo Next Captioner} checkpoint \cite{ghosh2026afnext},
while \emph{Qwen3-Omni} uses
\href{https://huggingface.co/Qwen/Qwen3-Omni-30B-A3B-Instruct}
{Qwen3-Omni-30B-A3B-Instruct} \cite{xu2025qwen3omni}.
Both are privileged captioning baselines with direct access to the
ground-truth target waveform, approximating an upper-bound captioning
setting without an additional caption-prediction stage. They use the
shared prompt in Fig.~\ref{fig:prompts}(b) and the official decoding
strategies. Their captions are used directly as CosyVoice3 style
instructions; all systems otherwise share the same TTS backend, response
text, and same-speaker prompt audio. These captions are not optimized
for CosyVoice3 and may therefore be suboptimal as control instructions.

\subsection{System-level results}
\label{sec:system-results}

Table~\ref{tab:main} shows that \model{} achieves the best performance
on all five downstream acoustic metrics in both partitions, despite the
privileged baselines having access to the ground-truth target audio.
Style instructions alone do not guarantee improvement: the unadapted
Base LLM obtains lower S2S scores and higher MCD-DTW than Raw TTS,
whereas speech-rewarded post-training reverses this degradation.

AF-Next provides a particularly clear example of the mismatch between
descriptive alignment and downstream control: it achieves the highest
T2S scores under all three speech--text encoders but ranks last on all
five downstream acoustic metrics in both partitions. This mirrors the
candidate-level mismatch observed in Sec.~\ref{sec:description-analysis}:
high descriptive alignment does not necessarily translate into effective
downstream control.

\subsection{LLM-based expressive speech evaluation}
\label{sec:preference}

\begin{table}[t]
\centering
\caption{Win rates (\%) in LLM-based expressive speech evaluation,
reported as \texttt{test\_in}/\texttt{test\_out}.
Rates exclude ties and invalid judgments
(together $<1\%$ of all judgments).}
\label{tab:preference}
\small
\setlength{\tabcolsep}{2pt}
\begin{tabular*}{\columnwidth}{@{\extracolsep{\fill}}lccc@{}}
\toprule
\multirow{2}{*}{System}
& \multicolumn{2}{c}{\model{} wins vs.\ system}
& System wins vs.\ GT \\
\cmidrule(lr){2-3}
& Context & Reference & Context \\
\midrule
Raw TTS    & 65.54/63.92 & 66.31/64.38 & 12.26/13.09 \\
Base LLM   & 54.19/53.67 & 60.91/57.75 & \underline{15.90}/16.87 \\
AF-Next    & 75.63/75.75 & 75.01/75.05 & 7.42/9.35 \\
Qwen3-Omni & 55.21/56.65 & 58.50/57.50 & 14.95/\underline{16.95} \\
\midrule
\model{} (ours) & -- & -- & \textbf{18.41}/\textbf{19.06} \\
\bottomrule
\end{tabular*}
\end{table}
A single recorded response represents only one possible realization of a reply, so reference consistency does not necessarily capture contextual appropriateness. We therefore complement target-reference evaluation with a context-only judgment that evaluates speaking style from the dialogue context without access to the target response audio. We use Gemini 3.8 Flash~\cite{gemini38flash} as an automatic audio judge in blind A/B comparisons on all utterances in both test partitions, with a low thinking level.  

In the context-only setting (Fig.~\ref{fig:prompts}(c)), the judge receives the three preceding utterances as audio with transcripts and the current response text, and selects the candidate whose speaking style is more appropriate for the reply. 
In the target-reference setting (Fig.~\ref{fig:prompts}(d)), it receives the context transcripts, current response text, and ground-truth response waveform, and selects the candidate whose style is more consistent with that waveform. Both candidates speak the same response text, and the judge is instructed to ignore speaker identity, voice timbre, and recording quality.

Table~\ref{tab:preference} shows that \model{} is preferred over all four baselines in both settings and partitions ($p<.001$, two-sided exact binomial tests). In the context-only setting, it achieves win rates of 54.19\%/53.67\% against the direct pre-RL Base LLM. Since the judge has no access to the target recording in this setting, this improvement indicates that speech-rewarded post-training also improves contextual appropriateness. In context-only comparisons with ground-truth speech, \model{} achieves the highest win rates among the synthesized systems (18.41\%/19.06\%). Ground-truth speech nevertheless remains preferred over all synthesized systems, highlighting the remaining gap in contextual appropriateness.

\section{Conclusion}
\label{sec:conclusion}
Our candidate-level analysis shows that speech--text alignment only weakly predicts downstream acoustic similarity. \method{} addresses this gap by training a conversational style planner with the teacher-forced target-speech likelihood of a frozen TTS model. The resulting planner outperforms all baselines on five downstream acoustic metrics across both evaluation partitions. LLM-based expressive speech evaluation further shows gains in contextual appropriateness and reference consistency. These results support optimizing natural-language instructions for their downstream synthesis behavior rather than relying solely on descriptive alignment. Our expressive-speech evaluation relies on an LLM judge and has not
been validated by human listening tests. In addition, both reward
computation and synthesis use CosyVoice3; evaluating transfer to other
TTS backends remains for future work.
\section*{Acknowledgment and AI Disclosure}
This work was supported in part by JST BOOST, Japan Grant Number JPMJBY25F6.
ChatGPT (OpenAI) was used to assist with language editing and phrasing
refinement throughout the manuscript, as well as with the development
of experimental scripts.

\bibliographystyle{IEEEbib}
\bibliography{refs}

\end{document}